\documentclass[twocolumn
]{revtex4-2}
\usepackage{amsmath}
\usepackage{caption}
\usepackage{graphicx}
\usepackage{latexsym}
\usepackage{dcolumn}
\usepackage{hyperref}
\usepackage{comment}
\usepackage{chemformula}
\usepackage{bm}
\usepackage{xcolor}

\begin{document}

% -------------------------------------------------------------------- %
% -------------------------------------------------------------------- %
% -------------------------------------------------------------------- %

% -------------------------------------------------------------------- %

% -------------------------------------------------------------------- %

% -------------------------------------------------------------------- %
\title{Kolmogorov–Arnold Chemical Reaction Neural Networks for learning pressure-dependent kinetic rate laws}

\author{Benjamin C. Koenig}
\affiliation{Department of Mechanical Engineering, Massachusetts Institute of Technology, 77 Massachusetts Ave, Cambridge, MA 02139, United States.}
\author{Sili Deng}
\email{Corresponding author, silideng@mit.edu}
\affiliation{Department of Mechanical Engineering, Massachusetts Institute of Technology, 77 Massachusetts Ave, Cambridge, MA 02139, United States.}

\date{\today}% It is always \today, today,

\begin{abstract} % 100 to 300 words.
Chemical Reaction Neural Networks (CRNNs) have emerged as an interpretable machine learning framework for discovering reaction kinetics directly from data, while strictly adhering to the Arrhenius and mass action laws. However, standard CRNNs cannot represent pressure-dependent or mixture-based rate behavior, which is critical in many combustion and chemical systems and {typically requires empirical falloff formulations such as Troe or SRI, or data-based interpolation or polynomial fits such as PLOG or Chebyshev Polynomials.} Here, we develop Kolmogorov–Arnold Chemical Reaction Neural Networks (KA-CRNNs) that generalize CRNNs by modeling each kinetic parameter as a learnable function of third-body concentrations using Kolmogorov–Arnold activations.
{This structure maintains the Arrhenius and mass action interpretability and physical constraints of a vanilla CRNN while enabling assumption-free inference of global and collider-specific pressure effects directly from data.} Two proof-of-concept reaction studies are presented to highlight the capability of KA-CRNNs to accurately reproduce pressure-dependent and collider-specific kinetics across a range of temperatures, pressures, and bath gas mixtures, extracting meaningful and generalizable models from sparse training data and significantly outperforming interpolative approaches (2.88x reduction in MSE). The framework establishes a foundation for data-driven discovery of extended kinetic behaviors in complex reacting systems, advancing interpretable and physics-{constrained} approaches for chemical model inference. 
\end{abstract}

\maketitle

\section{Introduction\label{sec:introduction}} \addvspace{10pt}

Chemical Reaction Neural Networks (CRNNs) are a recent machine learning tool for the autonomous inference of chemical pathways and kinetic parameters from time-series data \cite{ji_autonomous_2021}. They supplement traditional approaches that leverage substantial expert knowledge and quantum chemistry \cite{gao2016reaction} with an automated inference procedure to discover lean yet still expressive working models. Unlike standard black-box learning \cite{deng2025scientific, kumar2024combustion, koenig_multi-target_2023}, CRNNs guarantee reasonable model forms through a direct encoding of the Arrhenius and mass action laws in their network structures. The learned working models typically involve parameter counts on the order of 10 (as imposed by the sparse governing laws), compared to the thousands to millions typically present in deep neural networks. Their unique combination of analytical chemical laws and data-driven optimization enables template-free, semi-global working model inference from experimental data that can model visible trends and also capture intermediate species behavior.

Prior CRNN applications emphasize this unique position straddling data and theory-driven approaches. Early studies emphasized their autonomous capability to identify minimum sets of semi-global, ``working model'' pathways from limited TGA data, even with missing species \cite{ji_autonomous_2021, ji_autonomous_2022, li_bayesian_2023} and noisy measurements \cite{li_bayesian_2023, koenig_uncertain_2024}. Later studies incorporated physical priors to aid in convergence and improve the accuracy of the learned model to real chemical processes, via element conservation \cite{doppel_robust_2024} or enforcing known pathways in the CRNN structure \cite{koenig_accommodating_2023, koenig_comprehensive_2025}. Similar physics encoding using feedback loops with experimental work and theoretical computation enables further validation at the molecular scale \cite{wang2023kinetic, sun2024kinetic, chen2025hychem, koenig2026modeling}. Augmentations to include heat release rates inferred from DSC data \cite{koenig_accommodating_2023, bhatnagar_chemical_2025} and uncertainty quantification with TGA or DSC data \cite{li_bayesian_2023, koenig_uncertain_2024, koenig_comprehensive_2025} have further extended CRNNs to broader applications. Successful scientific and engineering applications include but are not limited to biomass pyrolysis \cite{ji_autonomous_2022, zhong2025pyrolysis}, energetic materials \cite{sun2024kinetic, wang2023kinetic}, enzymatic reactions \cite{ji_autonomous_2021, li_bayesian_2023}, heterogeneous reactions \cite{stagge2025findability, shukla2024discovering}, and lithium battery thermal runaway \cite{koenig_accommodating_2023, koenig_uncertain_2024, koenig_comprehensive_2025, bhatnagar_chemical_2025, koenig2026modeling}.

The direct Arrhenius and mass action law encoding is the major feature of CRNNs that enables this breadth of applications, but also limits their expressivity in more complex chemical rate dependencies. Many combustion reactions depend on the system pressure, for example, which traditionally requires additional analytical formulations (e.g., Lindemann \cite{lindemann1922discussion} or Troe \cite{gilbert1983theory} models) or interpolations (e.g., PLOG interpolation {or Chebyshev representations {\cite{venkatesh1997parameterization}}). Specific choices here have sizeable impacts on model expressivity, interpretability, and accuracy: for example, analytical formulations such as Troe or SRI provide conceptually interpretable theory-based falloff relations, yet with strict predefined mathematical formulas that do not necessarily represent real chemical behavior, for example in multi-well reactions. PLOG interpolations, meanwhile, retain physics-based Arrhenius formulas for the temperature dependnece and linearly interpolate in pressure, retaining temperature interpretation but abstracting the pressure dependence and potentially losing fidelity due to the log-linear interpolant. Chebyshev polynomials discard all presumed physics-based functions in favor of black-box fitting of the rate constant as a function of temperature and pressure, providing high accuracy and arbitrary curvature at a cost of little to zero interpretation of the learned expansion coefficients.} Further, mixture rules governing the combined effects of multiple colliders add additional complexity to the governing equations yet are a key source of uncertainty in state-of-the-art models \cite{singal2024implementation, burke2017evaluating, lei2020mixture}. A generalized CRNN approach capable of automatically inferring the rate effects of pressure, mixed bath gases, and other external effects would open the door to extended applications in combustion systems and beyond, but {clearly requires careful design in such a complex landscape of existing techniques and their tradeoffs}. One such option might involve the direct inference of parameterized empirical forms, such as the Troe model. Such an approach might involve updating the connections and activations within the CRNN to add the relevant model form with high-pressure, low-pressure, and falloff parameters.  However, this approach would limit the final accuracy to the assumptions present in the presumed model form, and limit algorithm application to regimes where such already-developed empirical models exist and are valid. PLOG-like interpolation between CRNN models is also feasible and enables learning without assumed model forms, although the logarithmic interpolation itself is an assumption on the pressure variation of the kinetic rates, introducing unnecessary error and limiting large-scale inference in an otherwise fully generalizable approach. {Of course, similarly to the Chebyshev polynomial approach, a standard MLP or PINN {\cite{raissi_physics-informed_2019}} solved via Neural ODEs {\cite{chen_neural_2019}} could represent the entire k(T, P) profile, but would provide zero interpretability or grounding in physical laws.} Thus, a desirable CRNN augmentation to capture pressure and mixture effects would retain the interpretability and accuracy of the underlying CRNN approach's kinetic dependence on temperature, while minimizing or eliminating the assumptions and simplifications present in standard pressure-based modeling approaches.

A promising recent machine learning tool showing potential for highly expressive and assumption-free inference while remaining visualizable and human-interpretable is the Kolmogorov-Arnold Network Ordinary Differential Equations (KAN-ODEs) approach \cite{koenig_kan-odes_2024}, which combines cutting-edge Kolmogorov-Arnold Networks (KANs) \cite{liu_kan_2024, koenig_leankan_2025} with Neural ODEs \cite{chen_neural_2019}. In it, KAN layers provide expressive and interpretable function representations via learnable activation functions (represented by weighted spline functions, similar to a finite element method), while ODE coupling enables direct inference of the rates of change in a dynamical system. In fact, KAN-ODEs have already proven their capability to accurately model complex chemical kinetic systems \cite{koenig_chemkans_2025}. In that study, however, much of the small-scale interpretability of individual KAN activations was sacrificed in order to develop a large enough model to capture all relevant chemical behaviors.

In this work, we propose a hybrid Kolmogorov-Arnold Chemical Reaction Neural Network (KA-CRNN) structure that assimilates these two techniques to fully leverage and maximize their unique benefits. The CRNN backbone retains exact adherence to the Arrhenius and mass action laws, providing strong inductive bias in the temperature dependence and facilitating a sparse overall structure. A relatively lean set of learnable KAN activations, meanwhile, defines each kinetic parameter as a unique, fully visualizable, and symbolically interpretable function of an external variable, here the pressure as parameterized by individual collider concentrations. This specifically designed insertion of KAN-based CRNN parameters enables assumption-free responses of the Arrhenius kinetic rates to the global pressure, collider-specific concentrations, and external factors, while retaining a highly sparse and fully interpretable chemical model representation. {Its dependence on fully parameterized Arrhenius and mass action temperature scalings coupled with a data-driven pressure function differentiate it from the Chebyshev polynomial approach {\cite{venkatesh1997parameterization}}, which involves a complete data-driven inference of the entire temperature and pressure response. This partial dependence on existing temperature rate forms enables visualization and human interpretation not available with the black-box Chebyshev polynomials: each Arrhenius parameter can be directly plotted as a function of collider concentrations, facilitating qualitative identification and understanding of kinetic trends.} This model development extends the scope of the CRNN model family to a broader range of chemical kinetic systems previously inaccessible to vanilla CRNNs. We demonstrate its functionality here in three proof-of-concept cases at varying temperatures, third-body concentrations, bath gas mixtures, pressure dependencies, and falloff functions.

\section{Methodology \label{sec:Methods}} 
\addvspace{10pt}

\subsection{Kolmogorov-Arnold Chemical Reaction Neural Networks}\label{sec:Methods_KACRNN}
\addvspace{10pt}

A generic Chemical Reaction Neural Network \cite{ji_autonomous_2021} models some $i^{th}$ reaction with two reactants (A, B) and two products (C, D),
\begin{equation} \label{eqn_reaction}
-\nu_{A,i} A + -\nu_{B,i} B \ch{->} \nu_{C,i} C + \nu_{D,i} D,
\end{equation}
where $A, B, C, D$ are the reactants and products and the corresponding $\nu_i$ are their stoichiometric coefficients, with a rate defined by the logarithmic form of the Arrhenius and mass action laws,
\begin{equation}
\begin{split} \label{eqn_rate_log}
r_i=\text{exp}(n_{A,i} ln[A] + n_{B,i} ln[B] \\ + ln A_i + b_i ln T - E_{a,i}/RT),
\end{split}
\end{equation}
\begin{equation}
\frac{d[X]}{dt}=\sum_{i} \nu_{X, i} r_i, \ \ \ X=A, B, ... \label{eqn_species_rate}
\end{equation}

Here $n_{A, i}$ and $n_{B, i}$ are the reaction orders, $E_{a, i}$ is the activation energy, $A_i$ is the frequency factor, and $b_i$ is the non-linear temperature factor. When wrapped into a neural network structure (a vanilla CRNN), the model can be solved forward, compared against training data, and then differentiated for iterative updates \cite{chen_neural_2019}.

{As discussed in the Introduction, a KA-CRNN requires the model to respond to the pressure as well. To do so here, we adapt the CRNN by first defining some external parameter $p_{ext}$ that we desire our model to respond to. Here, $p_{ext}=[M]$, the third-body concentration. The learnable kinetic parameters for some $i^{th}$ reaction are $KP_i = \{A, b, E_a\}_i$, where within a reaction we have $k=1:3$ parameters $KP_i^k$.} We encode each of these as simple univariate KAN activations \cite{liu_kan_2024, koenig_kan-odes_2024} defined per
\begin{equation} \label{eq:activ}
    KP_i^k = \text{KAN}\left(p_{ext}, \bm{\theta}_i^k\right) =
    {\phi}_i^k\left(p_{ext}\right),
\end{equation}
where KAN is a univariate function of $p_{ext}$ parameterized by the learnable $\bm{\theta}$. {In physical terms specific to this application, each distinct kinetic parameter $KP_i^k$ is encoded as a unique learnable univariate function of the third-body concentration $[M]$.} $\phi$ is the actual RBF formulation \cite{li_kolmogorov-arnold_2024} of the KAN activation,
\begin{align}
    \phi_{i} \left(\text{x} \right) &= \sum_{j=1}^{N} w^{\psi}_{i,j} \cdot \psi \left( \lvert \lvert \text{x}-c_{j} \rvert \rvert \right) + w^b_{i} \cdot b\left(\text{x}\right),\label{eq:basis}\\
    \psi(r)&=\exp(-\frac{r^{2}}{2h^{2}}). \label{eq:RBF}
\end{align}
In this formulation, $N$ is the KAN basis function grid size, and $w^{\psi}_{i,j}$ and $w^b_{j}$ are the learnable network parameters. These two sets of parameters encode the $j$ gridded basis function scales ($w^{\psi}_{i,j}$) and the single scale applied to the swish \cite{ramachandran_searching_2017} residual base function $b(\text{x})$ ($w^b_{i}$). Combined, they make up the learnable network parameter vector $\bm{\theta}$. The grid itself is defined by its uniform gridpoints $c_j$ and gridpoint spacing or RBF spreading parameter $h$. {As is standard for KAN-ODEs {\cite{koenig_kan-odes_2024}}, we normalize our grid to [-1, 1] and define the spreading parameter simply as the gridpoint spacing, $h=2/(N-1).$}

\begin{figure}
\centering
\includegraphics[width=0.9\linewidth]{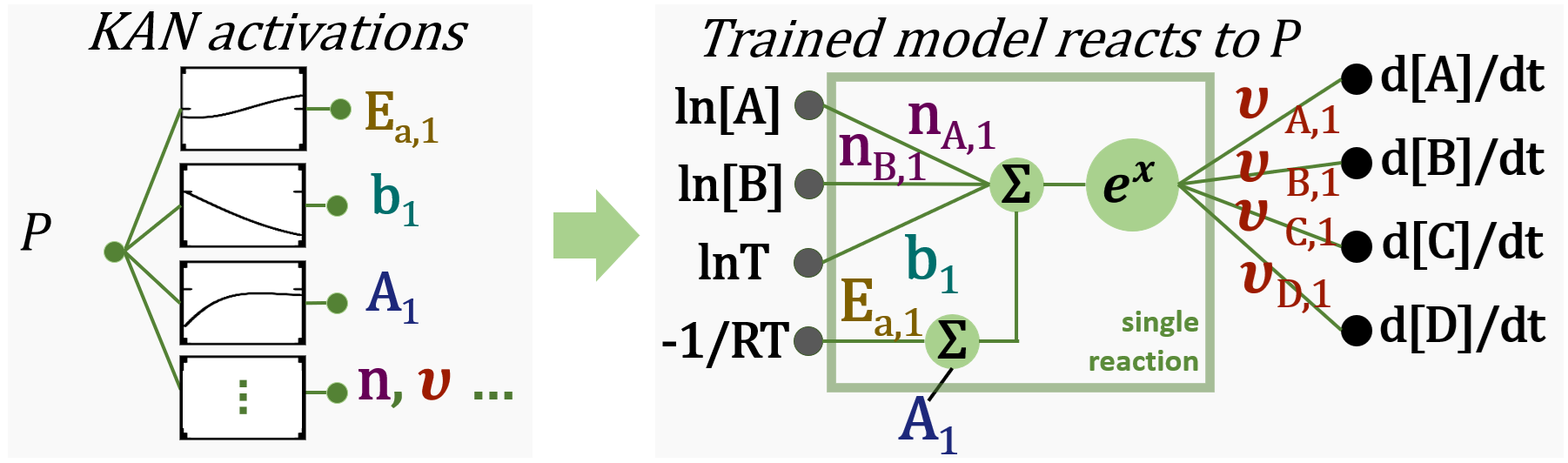}
	\caption{KA-CRNN with pressure-activated parameters.}
	\label{fig:fig1}
\end{figure}

The KA-CRNN redefines each kinetic parameter as its own univariate function of the pressure, as is visualized in Fig. \ref{fig:fig1}. Instead of the scalar parameter inference of a standard CRNN, inference in a KA-CRNN is carried out on the KAN parameters $\bm{\theta}$ that define the shape and magnitude of the continuous kinetic parameter functional responses to the pressure. Retained integration with the Arrhenius and mass action laws (as in a standard CRNN) keeps KA-CRNN activations simple, fully visualizable, and easily interpretable, compared to previous non-CRNN chemical kinetic applications of KAN-ODEs that required dozens of interconnected activations in series with low interpretability \cite{koenig_chemkans_2025}. {We finally remark that while the visualizable Arrhenius parameters inferred by the KA-CRNN provide significant human interpretability of kinetic trends with respect to pressure, there is no strict theory or hard constraint present in the modeling framework that guarantees the KA-CRNN finds the ``real'' form of the pressure dependence - a converged model is guaranteed only to fit the data and provide an interpretable model. That being said, we do provide comparison of the learned fits with the underlying Troe parameters, suggesting that the Arrhenius constraint on these models leads to a significant degree of chemically meaningful inferred behavior.}

The formulation presented in Eqs. \ref{eq:activ}-\ref{eq:RBF} can be applied to cases with an arbitrary number of colliders in the bath gas. For a single collider, as is studied in the first case here, $i=1:3$ to define a single set of KA-CRNN $E_a$, $A$, and $b$ parameters. For three colliders with different efficiencies or rate behaviors, meanwhile, we can simply extend to $i=1:9$ to define three unique sets of $E_a$, $A$, and $b$ for each of the three colliders.

Each forward pass of the KA-CRNN model is evolved using the DifferentialEquations package in Julia \cite{rackauckas_differentialequationsjl_2017}, with gradients on the KAN parameters $\bm{\theta}$ computed via autodifferentiation in the ForwardDiff package \cite{revels_forward-mode_2016} until convergence. The ADAM optimizer is used for learning \cite{kingma_adam_2017}, with the standard $1\times 10^{-3}$ learning rate. {The loss function is a simple MSE error based on the integrated species trajectories, with a heavy PINN constraint on negative activation energies $E_a$ and pre-exponential factors ln$A$ to ensure these parameters stay non-negative.} To ensure convergence, the KA-CRNN activations are initialized as the constant values defining the high-pressure kinetics (inferred via standard CRNN on a single high-pressure dataset). {Complete training details are available in the Supplementary Materials.}

\section{{Illustrative Case Studies}}

{Three case studies are presented in this work. Their motivations, formulations, and intended demonstrations of inference capabilites are outlined here. A summary of the temperature, pressure, and initial conditions is available in the Supplementary Material. The ground-truth models used to generate training data are available in the Supplementary Material.}

\subsection{Pressure-sensitive reaction model\label{sec:data} }
 \addvspace{10pt}

In a preliminary case designed to demonstrate the capability of KA-CRNNs to learn parameters as continuous functions, we investigate the pressure-dependent single-step CH$_3$ recombination rate discussed in \cite{wagner1988study} and used as a canonical example in the Chemkin theory manual \cite{ansys}. The ground truth rates are computed using the Troe formulation \cite{gilbert1983theory}, which incorporates nine equations and nine parameters to fully describe the falloff behavior. We assume a homogeneous bath gas of unity efficiency and compute isothermal trajectories from an initial concentration of $10^{-11}$ mol/cm$^3$. In this case, we remark that the low-pressure and high-pressure rate constants used in the true governing equations follow the general form $k = A T^b \text{exp}(-E_a/RT),$ aligning mathematically with the formulation used in Eq. \ref{eqn_rate_log} of the KA-CRNN. This alignment allows us to isolate testing of the pressure-sensitive learning capability in an otherwise straightforward setting.

\subsection{Linear mixture rule formulation \label{data_2}} 
 \addvspace{10pt}

In a second case, we highlight the KA-CRNN's robustness by learning collider-specific rate parameterizations from training data with mixtures of bath gases, for which the governing law rate constants and KA-CRNN rate constants take different functional forms. To this aim, we study the $H + O_2 (+M) \ch{->} HO_2(+M)$ combination reaction in various mixtures of Ar, N$_2$, and H$_2$O. The rate parameterizations \cite{baulch2005evaluated} from which training data are generated include a universal $k_\infty$ of the form $k_\infty=A_1 T^{b_1} + A_2T^{b_2}$, three collider-specific $k_0$ parameterizations of the form $k_0 = AT^{b}$, and finally three unique scalar values for the center broadening factors $F_c$. Combined with the Troe formulation \cite{gilbert1983theory}, these parameterizations enable the computation of rate constants $k_i$ for each of the three colliders $i$ at arbitrary concentrations. Then, to represent the overall reaction rate, we use a mole fraction-weighted average of the three rate constants as per the linear mixture rule,

\begin{equation}
    k^{eff} = \Sigma_{i=1}^3k_i x_i. \label{eq:mix_rule}
\end{equation}

As in the first case study, we generate isothermal trajectories at varying temperatures from an initial concentration of $10^{-11}$ mol/cm$^3$. Here, however, we vary the mixture of bath gases at each sample as will be discussed in Sec. \ref{sec:results_2}. Unlike the first case, we remark here that the $k_0$ and $k_\infty$ parameterizations do not share a model form with the presumed [$E_a, A, b$] kinetic parameterization used in the KA-CRNN. Thus, in this case, we test both the KA-CRNN's capability to extract distinct collider parameterizations from mixed gas data, as well as its ability to capture data that does not neatly conform to its Arrhenius-based parameterization.

\subsection{Noisy SRI data with random initializations}

{The two preceding cases studied the KA-CRNN's capability to simultaneously reconstruct idealized synthetic data and falloff plots with high accuracy, in increasingly complex toy conditions. Here, we aim to test the robustness of the methodology to increasingly realistic considerations.

First, we test the KA-CRNN's structure-invariant convergence by moving from the Troe falloff function (used in both preceding cases) to the five-parameter SRI form} \cite{stewart1989pressure, kee1989chemkin}. {Second, we select the chemically activated bimolecular reaction $CH_3 + CH_3 (+M) \ch{->} C_2H_5 + H (+M)$, which has an inverse pressure dependence (decreasing rate as a function of [M]). The basic three-parameter SRI falloff parameters for this reaction} \cite{stewart1989pressure, ansys} {are numerically augmented here with the assignment of $d=0.9$ and $e=-0.1$, where $d=1$ and $e=0$ is identical to the three-parameter form. This modification is carried out to increase the numerical complexity of the system for demonstration purposes, and does not necessarily reflect real chemical behavior. Third, we add 5$\%$ additive synthetic noise to the training data, the magnitude of which is evaluated on the maximum value and then applied uniformly (and randomly) to all datapoints. Fourth and finally, we repeat the inference process ten times: five times with random KAN activation initializations in the high-pressure range (as inferred by a vanilla CRNN on the highest-pressure dataset), and five times with random KAN activation initializations at the low-pressure limit (as inferred by a vanilla CRNN on the lowest-pressure dataset). The random initializations are defined with a 10$\%$ noise around the converged vanilla CRNN values. 

Overall, with this case we aim to move closer to a realistic experimental setting and demonstrate the robustness of KA-CRNNs to structural differences in falloff forms, noisy data, and initialization ablation.}

\section{Results and Discussion} \label{results}
 \addvspace{10pt}
\subsection{Pressure-sensitive reaction model \label{results_1}}

In the first case study, we provide a preliminary test of the KA-CRNN algorithm's ability to simultaneously infer pressure and temperature-sensitive kinetics in an otherwise straightforward learning problem. The KA-CRNN is trained to learn a single set of pressure-sensitive $E_a$, $A$, and $b$ values that reconstructs the reaction detailed in Sec. \ref{sec:data} across a gridded set of 27 conditions: temperatures of \{800 K, 1000 K, 1200 K\}, and third-body concentrations of $10^{-i}$ mol/cm$^3$ for $i=2:10$ (including the falloff region). Each activation contains only eight gridpoints and one base activation, leading to an extremely sparse KA-CRNN with just 27 parameters representing the entire model. Reaction orders and stoichiometric coefficients are fixed at their known values. Normalized third-body concentrations are used as the effective pressure input in the KAN activations of Eq. \ref{eq:activ}.

\begin{figure}
\centering
\includegraphics[width=\linewidth]{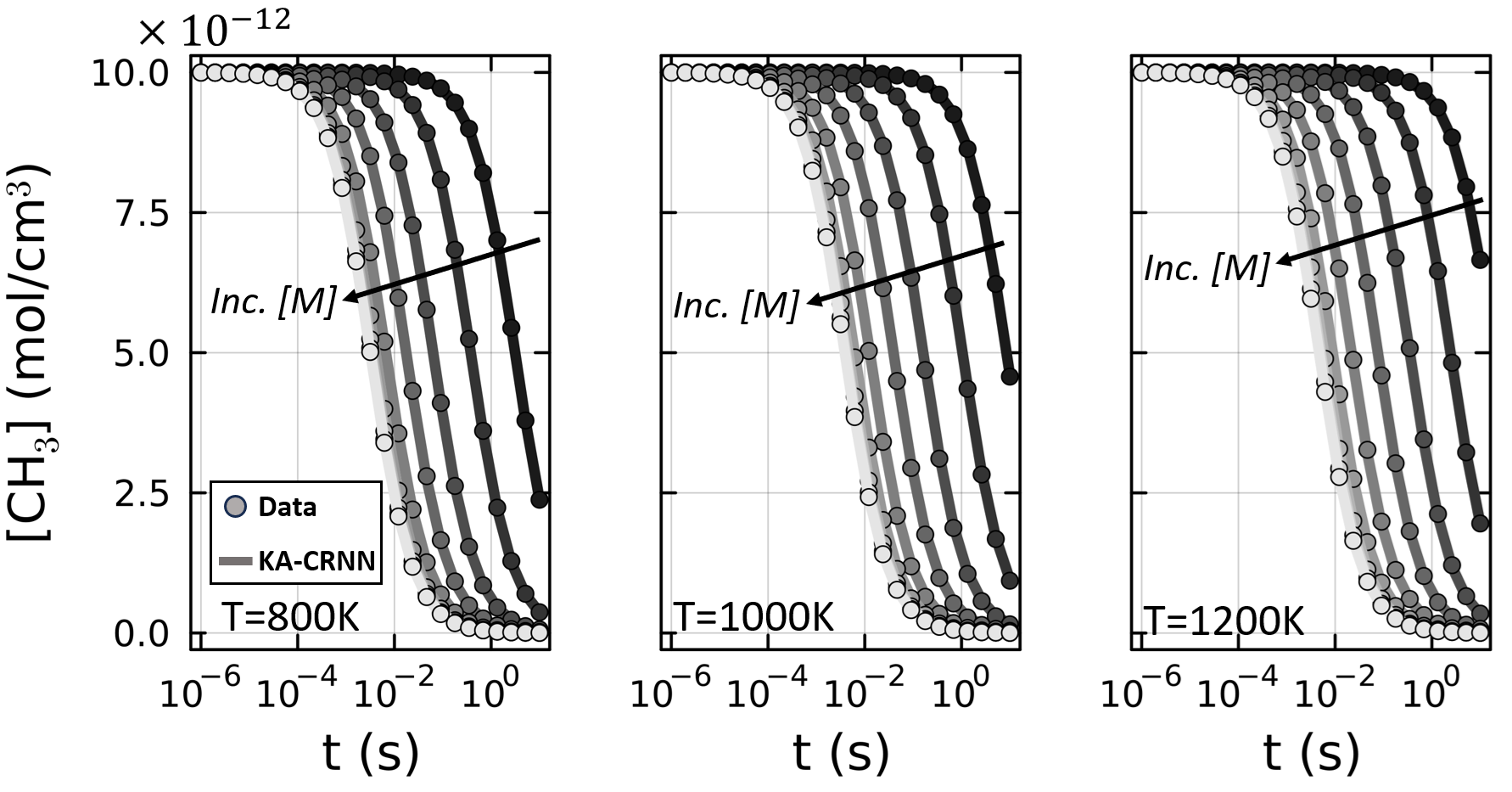}
	\caption{KA-CRNN reconstructions of 27 single-collider datasets using pressure-sensitive $E_a$, $b$, and ln$A$. Profiles evaluated at [M] = $10^{-i}$ mol/cm$^3$, $i=2:10$, at three temperatures labeled on subplots.}
	\label{fig:recon_species}
\end{figure}

KA-CRNN reconstructions for all 27 training profiles are shown in Fig. \ref{fig:recon_species}, where we observe strong agreement at all temperatures and third-body concentrations (pressures). The KAN activations encode the model's sensitivity to the pressure, while the underlying CRNN's direct exploitation of the Arrhenius law encodes its sensitivity to the system temperature, enabling the compact model to accurately reconstruct all datasets. The converged mean squared error loss here was $2 \times 10^{-16}$. {As a benchmark, a standard black-box multi-layer perceptron network was trained to directly compute $k = MLP(T, [M])$ with no CRNN encoding and no KAN parameters. Otherwise, the training problem was identical between the KA-CRNN and MLP: from an initial guess, the MLP was integrated forward in time to reconstruct the training data and then iteratively updated to minimize reconstruction error. A larger 177 black-box parameters was found necessary for proper convergence, with a converged accuracy of $1 \times 10^{-15}$. Both of these MSE loss values are quantitatively very low, and are in fact nearly indistinguishable if plotted together as in Fig. {\ref{fig:recon_species}}. We retain this quantitative discussion to highlight that the physical interpretation and lean parameterization of the KA-CRNN do not lead to any MSE cost, as a traditional regularizer might: it takes far greater MLP parameters to converge, and even further to converge to a similar accuracy.}
Further detailed discussion of the quantitative benefits of CRNNs and KAN-ODEs over MLPs is available in previous work \cite{li_bayesian_2023, koenig_kan-odes_2024, koenig_chemkans_2025}.

\begin{figure}
\centering
\includegraphics[width=\linewidth]{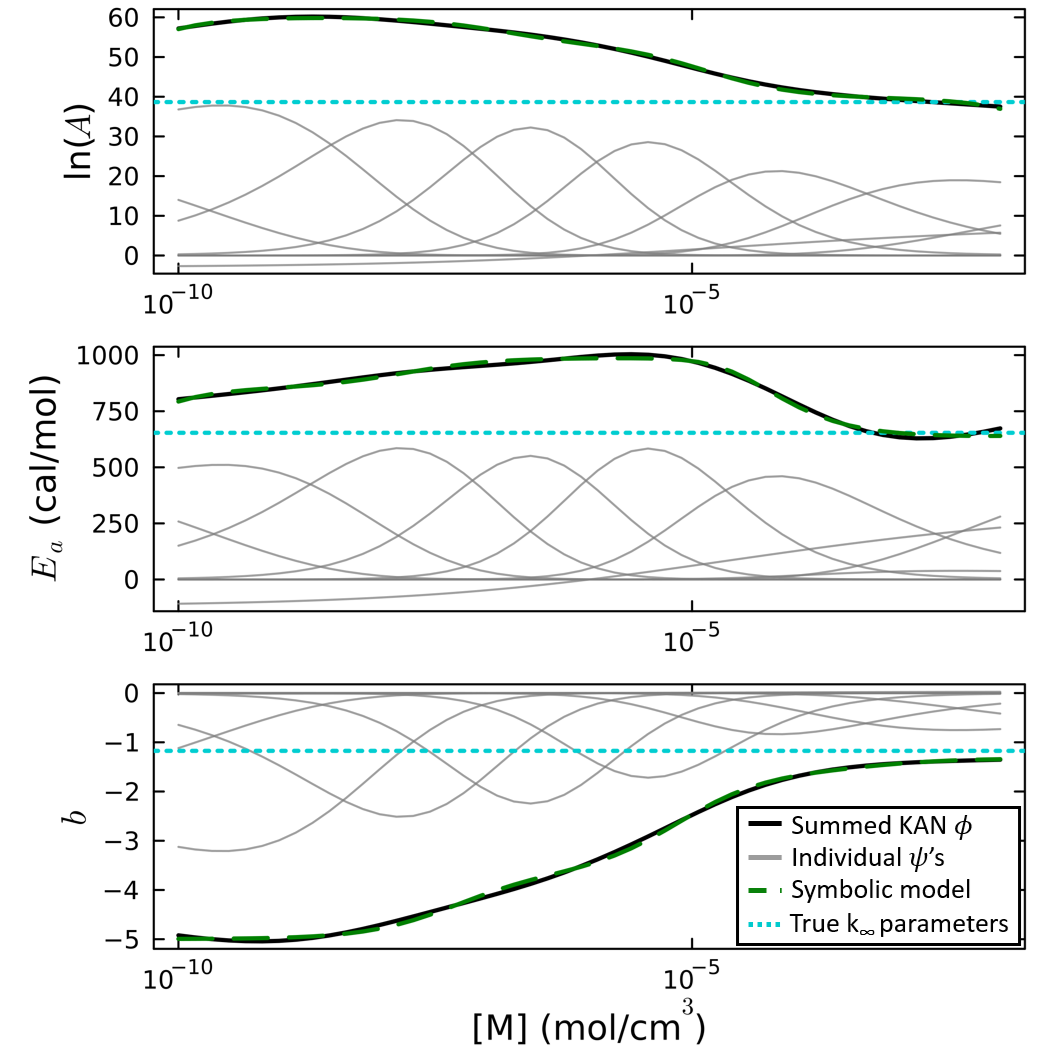}
	\caption{KAN activations and individual basis functions of kinetic parameters in the single-collider Troe case. Nine network parameters used for each activation as shown by light grey traces ($9 \times 3 = 27$ total). True Troe $k_{\infty}$ values shown in cyan, which the KA-CRNN automatically reconstructs well at high pressure.}
	\label{fig:KAN_acts}
\end{figure}

The KAN activations of the three learned kinetic parameters are visualized in Fig. \ref{fig:KAN_acts}. Thicker black traces show the overall parameter responses to the third-body concentration (pressure), while the lighter gray traces visualize the basis functions that sum to recreate each parameter. KAN models benefit from such interpretability and visualizability during and after the training process, especially in the specialized univariate KA-CRNN implementation, where each kinetic parameter uses a single activation only. {Also plotted in Fig. {\ref{fig:KAN_acts}} in cyan are the three kinetic parameters making up the ``ground truth'' Troe model's constant $k_{\infty}$, which the KA-CRNN is seen to converge to at the high-pressure limit.} A further capability of KAN activations is their propensity for symbolic regression \cite{liu_kan_2024, koenig_kan-odes_2024}. Here, we postprocess these activations using the SymbolicRegression.jl package \cite{cranmerInterpretableMachineLearning2023} with the four fundamental binary operators $[+,-,\times,/]$ and three unary operators $[sin(), cos(), exp()]$. The {application-ready symbolic formulas}, which are also plotted in green in Fig. \ref{fig:KAN_acts}, are 

\begin{align}
&\text{ln}A = \phi_1([M]) = \frac{7.30}{7.43\text{E}6 [M] + 1.05}- 272.8 [M] \\
& \ \ \ +\frac{1.91\text{E-}4}{[M] + 1.42\text{E-}5}  + 39.66 - \frac{3.03\text{E-}10}{[M]}, \notag \\
&E_a = \phi_2([M]) = -353.5 \left( e^{-\frac{5.52\text{E-}5}{[M] + 7.10\text{E-}6}} - [M] \right)\\
& \ \ \  - \frac{2.12\text{E-}6}{[M] + 1.58\text{E-}8} - \frac{6.13\text{E-}9}{[M]} + 987.9, \notag \\
&b = \phi_3([M]) = \frac{[M]}{4.04 [M] + 0.00240}  \\
& \ \ \ - \frac{1.55\text{E-}5}{[M] + 7.43\text{E-}6} - \frac{6.51\text{E-}8}{[M] + 4.89\text{E-}8} - 1.58. \notag
\end{align}
{While not trivially human-interpretable like the activations shown in Fig. {\ref{fig:KAN_acts}}, these symbolic formulas provide easily computed kinetic parameters for cases where handling the 27-term, rbf-based KAN activations may not be preferred. Discrete, pre-computed look-up tables are an additional method to simplify these activations before use in iterative solvers or other computationally intensive tasks.}
\begin{figure}
\centering
\includegraphics[width=\linewidth]{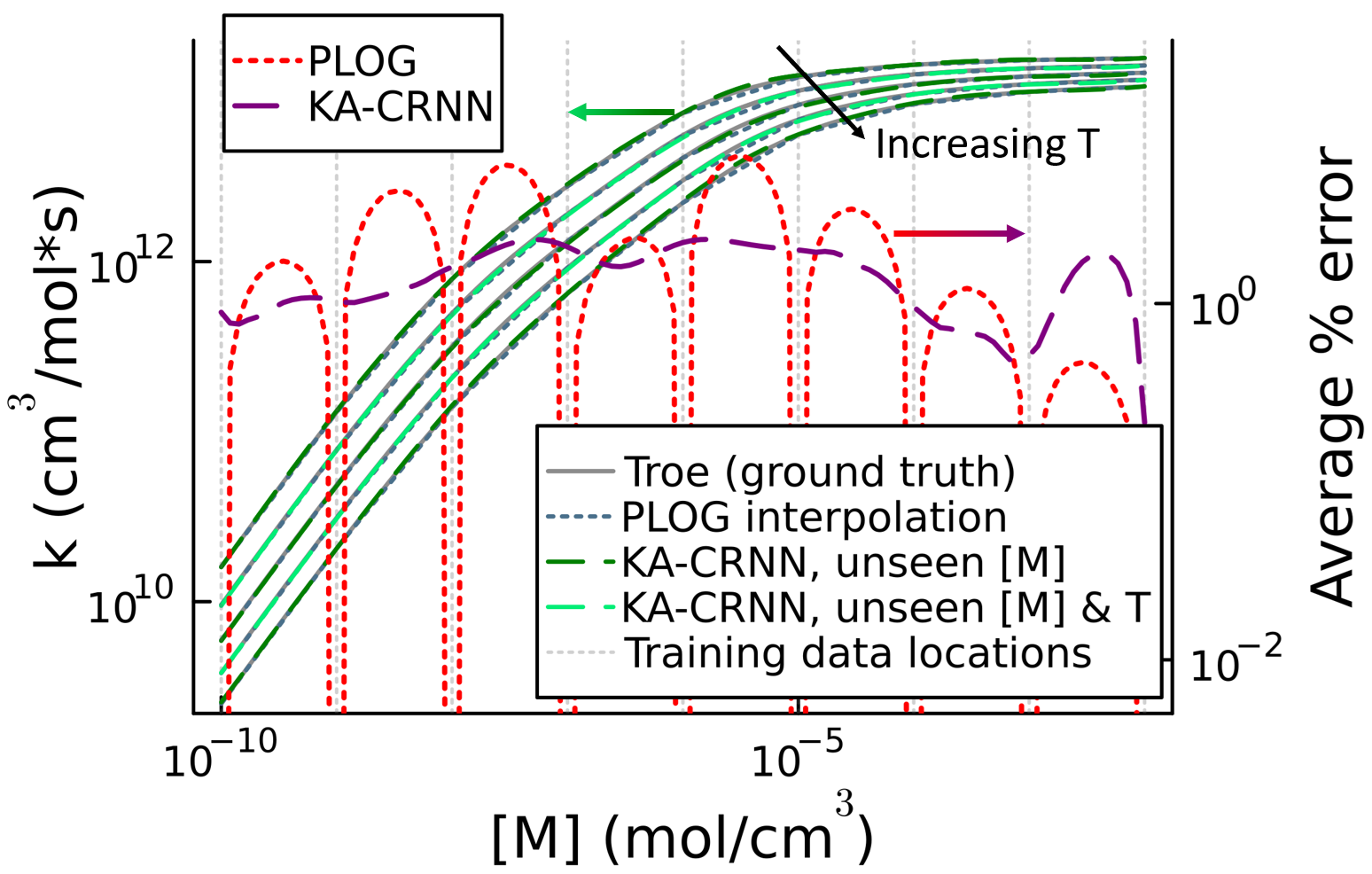}
	\caption{k vs. [M] inferred by KA-CRNN (left axis). Synthetic ``ground truth'' from Troe model. Discrete training data locations shown with vertical dashed lines. Unseen testing at all intermediate [M] values  and at {900 K, 1100 K} (light green traces). Average percent error (right axis) shows low KA-CRNN error throughout, while PLOG rapidly accumulates error due to loglinear shape.}
	\label{fig:k_m}
\end{figure}

Finally, we present KA-CRNN generalization results for entirely unseen testing conditions. The training data contains continuous species profiles in time at 27 discrete (T, [M]) combinations. In Fig. \ref{fig:k_m}, we generalize [M] to the continuous values between the training points, and we generalize T to discrete values unseen during the training cycle. The KA-CRNN is able to generalize to the unseen intermediate [M] and T values very well, as is evident in the smooth and highly accurate reaction rate reconstructions for all training and testing curves. To benchmark these results, we compare against the PLOG interpolations on the same reaction. We provided the PLOG interpolants, which do not have the rate inference capabilities of CRNN or KA-CRNN, with the optimal ``ground truth'' Troe parameters at all 45 data locations: 9 {third-body concentrations} and 5 {temperatures} (3 training and 2 testing). This enabled fair comparison in a best-case scenario for the PLOG solution, while the KA-CRNN was tasked with both inferring the parameters and also interpolating in pressure. {The rate constants themselves, plotted on the left hand axis, are not easily distinguishable. We have additionally plotted the average percent error (at all five temperatures) on the right axis. As expected, PLOG beats the KA-CRNN at the exact training [M] locations, as the PLOG interpolant is given the $0\%$ error ground truth at those points. As also expected, however, the PLOG interpolation degrades significantly when not in the immediate vicinity of these ground truth locations, while the KA-CRNN maintains a smooth error profile near {1$\%$} throughout the entire domain.} The MSE in rate constant reconstructions from the KA-CRNN at all five temperatures is 2.88 times smaller than that of the PLOG interpolations, despite the PLOG being provided with these ground truth models. This significant degree of accurate KA-CRNN inference and generalization is made possible by the extremely sparse and overfitting-resistant KAN activations \cite{koenig_chemkans_2025}, as well as the direct enforcement of the Arrhenius and mass action laws from CRNN.

\subsection{Linear mixture rule formulation \label{sec:results_2}} 

\begin{figure*}[!t]
\centering
\includegraphics[width=.92\linewidth]{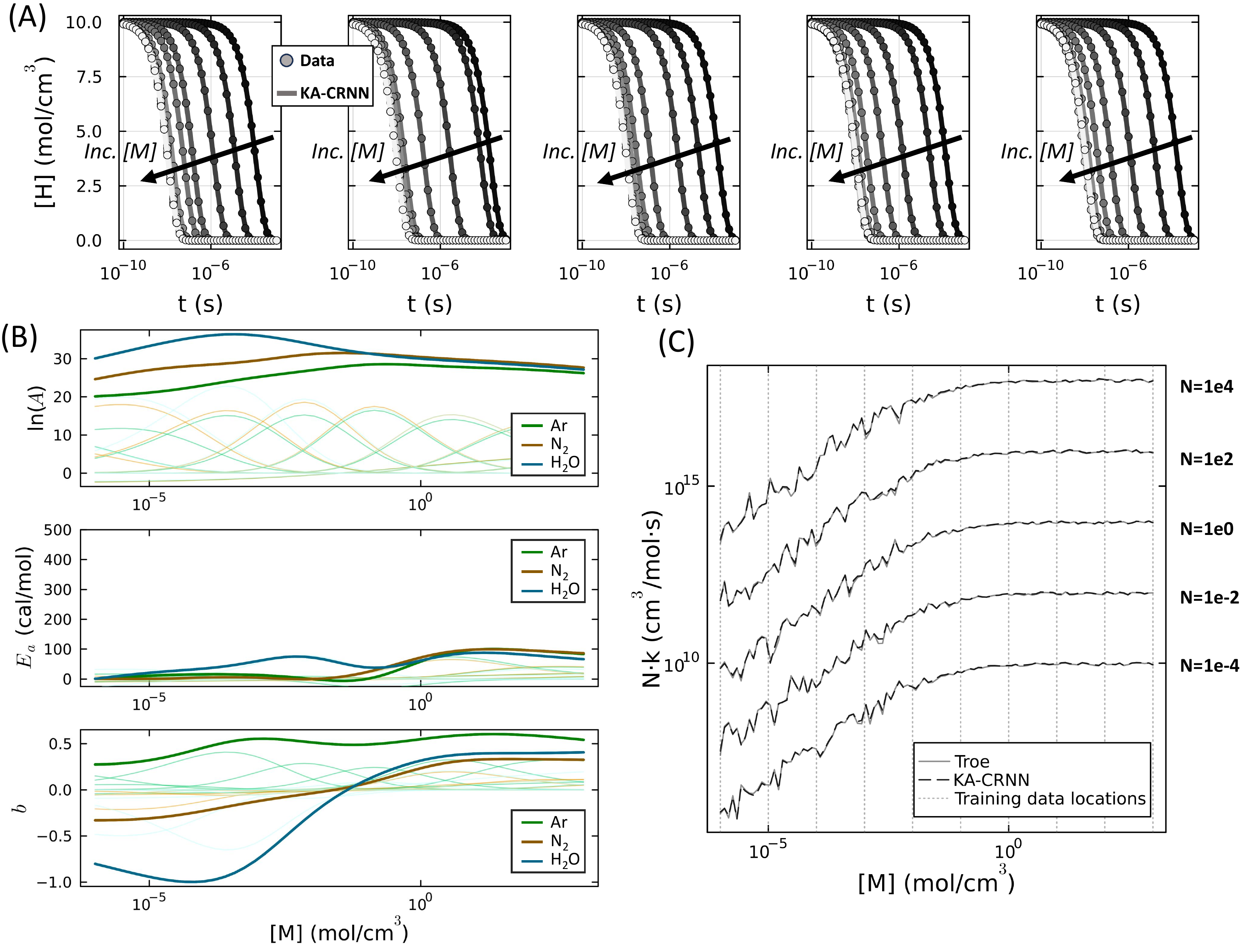}
	\caption{Results for mixture rules case. (A) Reconstructed training data at [M] = $10^{i}$ mol/cm$^3$, $i=-6:3$. Each curve randomly samples the bath gas mixture ratio and T within [800K, 1200K] (each [M] sampled once per subfigure). (B) KA-CRNN kinetic parameters for each collider. Trends across low and high pressures reflect Troe ground truth, where the three colliders have distinct low-pressure rate constant formulations but converge to a single high-pressure rate constant. (C) Validation at unseen intermediate [M] values and unseen random mixture ratio and T samples (repeated 5 times, with vertical offset for visualization). }
	\label{fig:mixture}
\end{figure*}

In the second case study, we increase the complexity of the learning problem and move closer to practical applications through an investigation of the KA-CRNN's capability to extract collider-specific kinetic relationships from global data of mixed bath gases. To this aim, we retain a structured sampling grid of third-body concentrations, using [$10^{i}$] mol/cm$^3$ for $i = -6:3$ (adjusted from the first case to reflect this reaction's falloff range). Here each third-body concentration value is randomly sampled five times, where a given random sample includes both a uniform random temperature in the range of [800 K, 1200 K], as well as a random ratio of the three bath gases. We note that while the precise ratios of the gas blends are randomly sampled, the absolute third-body concentration [M] remains on the structured grid. Combined, there are thus 50 total time series profiles used to train the KA-CRNN. Reaction orders and stoichiometry are again fixed at their known integer values. The overall task here is for the KA-CRNN to not only fit the data, but also to extract distinct and accurate representations for the three colliders' kinetic rates that enable full generalization to unseen intermediate concentration values, unseen random temperature samples, and unseen random bath gas mixtures. 

A KA-CRNN kinetic parameterization ($A, E_a, b$) is learned for each collider using 27 parameters, as in the first case study. The linear mixture rule of Eq. \ref{eq:mix_rule} is directly encoded, where the overall rate constant integrated forward is the mole fraction-weighted sum of the three collider-specific, KA-CRNN computed rate constants. While higher-fidelity mixture rules {such as LMR-R} exist \cite{burke2017evaluating}, we retain the linear formulation in this proof-of-concept as it remains widely used in many academic and industrial contexts. We remark that the implementation of the mixture rule occurs in series with the KA-CRNN but remains a fully distinct and separable step, making generalization to more complex or nonlinear mixture rules mathematically straightforward.

KA-CRNN reconstructions of the data are shown in Fig. \ref{fig:mixture}(A). There, each individual subplot contains a complete set of ten sampled [M] values, with randomized temperatures and gas ratios at each of the ten samples. This sampling process is repeated five times total (once per subplot). As the sampled [M] increases by 10x for each location on the structured grid, the corresponding lines and markers in Fig. \ref{fig:mixture}(A) move toward earlier times with increased [M] (and correspondingly lightened coloring), thanks to the increase in rate constant as a function of [M]. The KA-CRNN is seen to fit all data well, which is an early signal of its capability to capture mixture effects and to automatically resolve model form discrepancies. We remark as well that as the high-temperature rate constant approaches the global maximum $k_\infty$ at high [M] values, variations in temperature become more evident: in the majority of Fig. \ref{fig:mixture}(A)'s subplots, the fastest trajectory was not that at the highest [M], but that in the high-pressure regime with the highest random temperature sample, as seen for example in the third and fourth subplots where darker grey markers indicating lower [M] values are seen to react slightly faster than the lightest grey markers. The KA-CRNN's accurate modeling of this behavior signals its ability to balance learning not just the high-variance low-pressure and falloff region effects, but also the relatively [M]-invariant high-pressure regime where minute rate differences are primarily impacted by standard temperature-driven kinetics.

We additionally plot the collider-specific KA-CRNN activations in Fig. \ref{fig:mixture}(B). There, the lighter curves are the individual KAN basis functions and base activations, while the three darker curves per subfigure are the summed pressure-dependent Arrhenius parameters. While the training profiles contained only the global [H] trajectories in Fig. \ref{fig:mixture}(A), we see in Fig. \ref{fig:mixture}(B) that the KA-CRNN was able to extract from this global information the distinct effects of the three colliders. As expected from the underlying theory, all three parameters for the three colliders generally tend to converge toward higher [M] values, while at lower values, there is more distinct behavior reflecting the transition to the falloff region and low-pressure regime. Aside from negative activation energies in the low-pressure regime for $H_2O$, we also interestingly note that the activation energies were largely driven toward zero, as might be expected from the the non-exponential temperature dependences used in the governing rate laws \cite{baulch2005evaluated}.

In most CRNN applications, overfitting is not observed thanks to the extremely sparse network parameterizations (a result of the direct encoding of the real governing laws) and the correspondingly large data to parameters ratio. Here, however, we have run a fairly significant validation study in Fig. \ref{fig:mixture}(C) to ensure that the addition of the KAN activations has not negatively affected the CRNN's generalization capability. We computed rate constants (after applying the mixture rule) on a denser, 100-point [M] grid between the minimum and maximum values used in training, where we recall that the training data was sampled on a much sparser 10-point grid. Thus, as in Sec. \ref{results_1} the first capability demonstrated here is the KA-CRNN's interpolation to unseen intermediate third-body concentrations. To add further unseen behaviors to the validation set, we additionally generated a new random temperature and bath gas mixture at each of those 100 [M] locations, extending the unseen generalization to the temperature and mixture spaces as well. This random sampling process was repeated five times, with all rate constant results plotted in Fig. \ref{fig:mixture}(C) with a vertical offset to aid in visualization. Essentially, the validation set takes the rather sparse 50-point sample in the space of third-body concentrations, temperatures, and mixture ratios and then rigorously tests if it can fully generalize to all potential intermediate conditions. The near-perfect rate constant reconstructions shown throughout Fig. \ref{fig:mixture}(C) confirm the KA-CRNN's generalization capability: its strong match to the jagged randomized Troe rate constant profiles demonstrates that it handles all relevant pressure, mixture, and temperature-dependent effects with high accuracy, showing no perceptible performance degradation at the unseen conditions. {This is of particular note in the current case given the structural mismatch between the two-term ``ground truth'' Troe rate constant and the KA-CRNN's presumed single-term approximation: while we anticipate this mismatch to lead to a certain amount of systematic error, we see in these results that the data-driven nature of the KA-CRNN is still able to fit the overall rates and reconstruct the training and testing data well. Further, in fact, Fig. {\ref{fig:mixture}(B)} exhibits two kinetic parameter behaviors: low-pressure variation in values across the three colliders, and high-pressure convergence across all to the same values. This quite accurately reflects the underlying Troe form, in which the three colliders share a $k_{\infty}$ formulation but each have distinct $k_0$ formulations. When cross-referenced against Fig. {\ref{fig:mixture}(C)}, the visualizable kinetic parameters of Fig. {\ref{fig:mixture}(B)} are further all seen to exhibit a regime shift from low-pressure, well-distributed behavior to high-pressure, converged behavior in the same [M] range as the main falloff behavior in the underlying models. Thus, even when strong prior knowledge of the exact nature of a reaction's temperature dependence is not known, the KA-CRNN remains able to fit a hybrid theory/data-driven model that captures the relevant trends in temperature, pressure, and collider-specific adjustments while also revealing key insights surrounding these trends through its interpretable kinetic parameter activations.}

In summary, we have demonstrated in this section that while retaining highly sparse network parameterizations (81 learnable parameters composing nine [M]-sensitive kinetic model parameters), the KA-CRNN approach can effectively learn collider-specific kinetic rate models under an array of confounding variations in temperature, third-body concentration, and bath gas mixture ratio. These proof-of-concept studies extend the applicability of the underlying CRNN framework to a broader set of chemical kinetic situations where unknown or poorly-understood responses to non-Arrhenius inputs play a substantial role, and demonstrate the KA-CRNN's positioning to help further refine known model pathways and parameterizations or to generate new global pathways purely from time-series data.

\subsection{{SRI model with noisy data and randomized initialization}}

\begin{figure}
\centering
\includegraphics[width=0.8\linewidth]{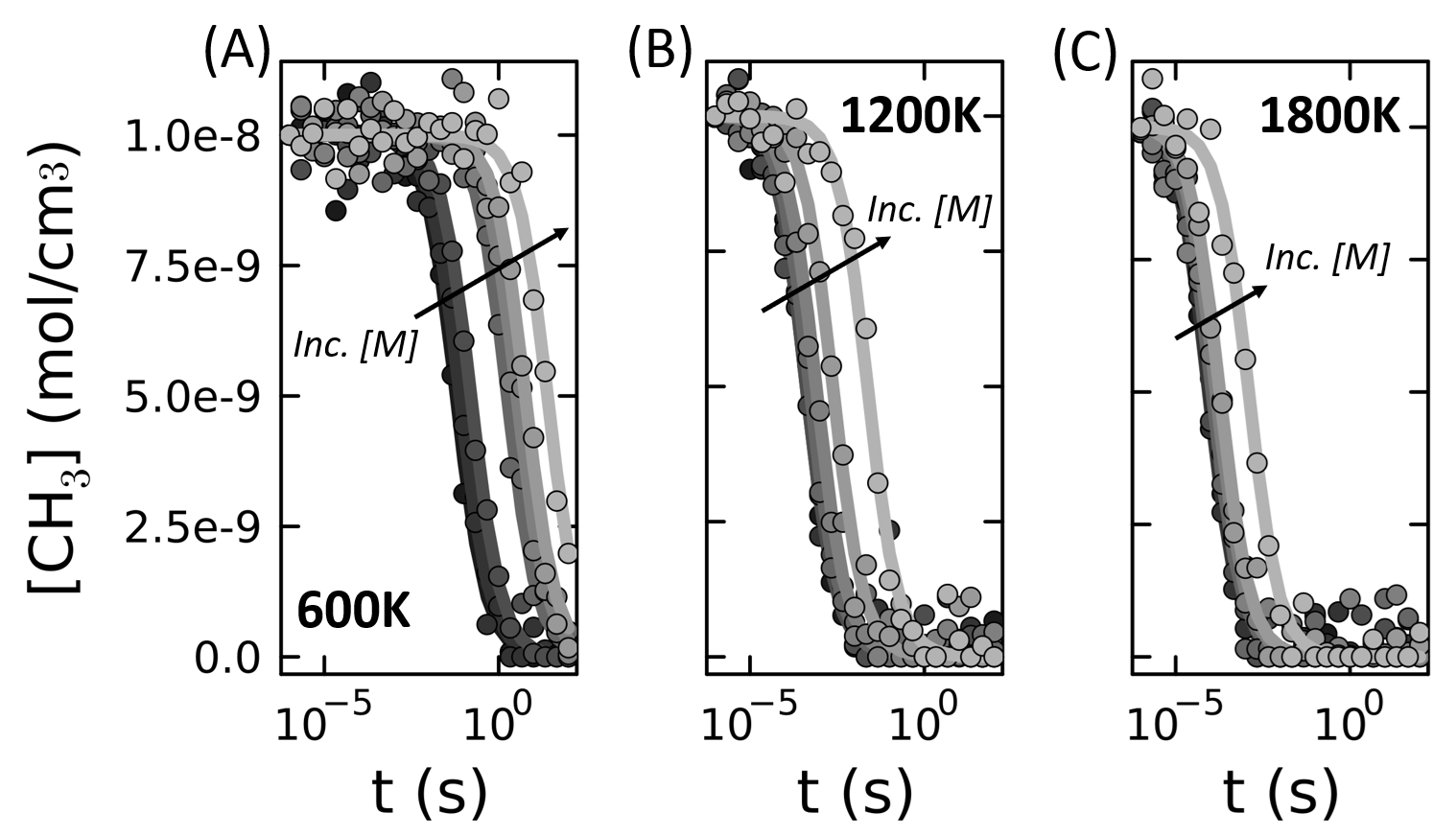}
	\caption{{Converged KA-CRNN model (curves) on noisy SRI falloff data (dots). Ten cases studied in this section - just one plotted here for visualization. Increasing [M] values are on a 10x interval between $10^{-7}$ and $10^{-2}$.}
    }
	\label{fig:sec3_recon}
\end{figure}

{We finally present brief results highlighting the robustness of KA-CRNNs to experimental noise, structural variation in the underlying falloff function, the direction of the pressure dependence, and randomness in the initializations. Fig. {\ref{fig:sec3_recon}} shows the converged KA-CRNN fits for one high-pressure initialization case. Note the direction of the [M] dependence has reversed compared to Figs. {\ref{fig:recon_species}} and {\ref{fig:mixture}(A)}. Additionally, we note a stronger positive correlation of reaction rate and temperature than was the case in the two preceding studies. The KA-CRNN demonstrates robustness to the noise, both in its accurate inference of the main temporal dynamics of the reaction and in its accuracy in carrying the reaction to completion, even with significant noise in the near-zero concentration data.}

\begin{figure}
\centering
\includegraphics[width=0.74\linewidth]{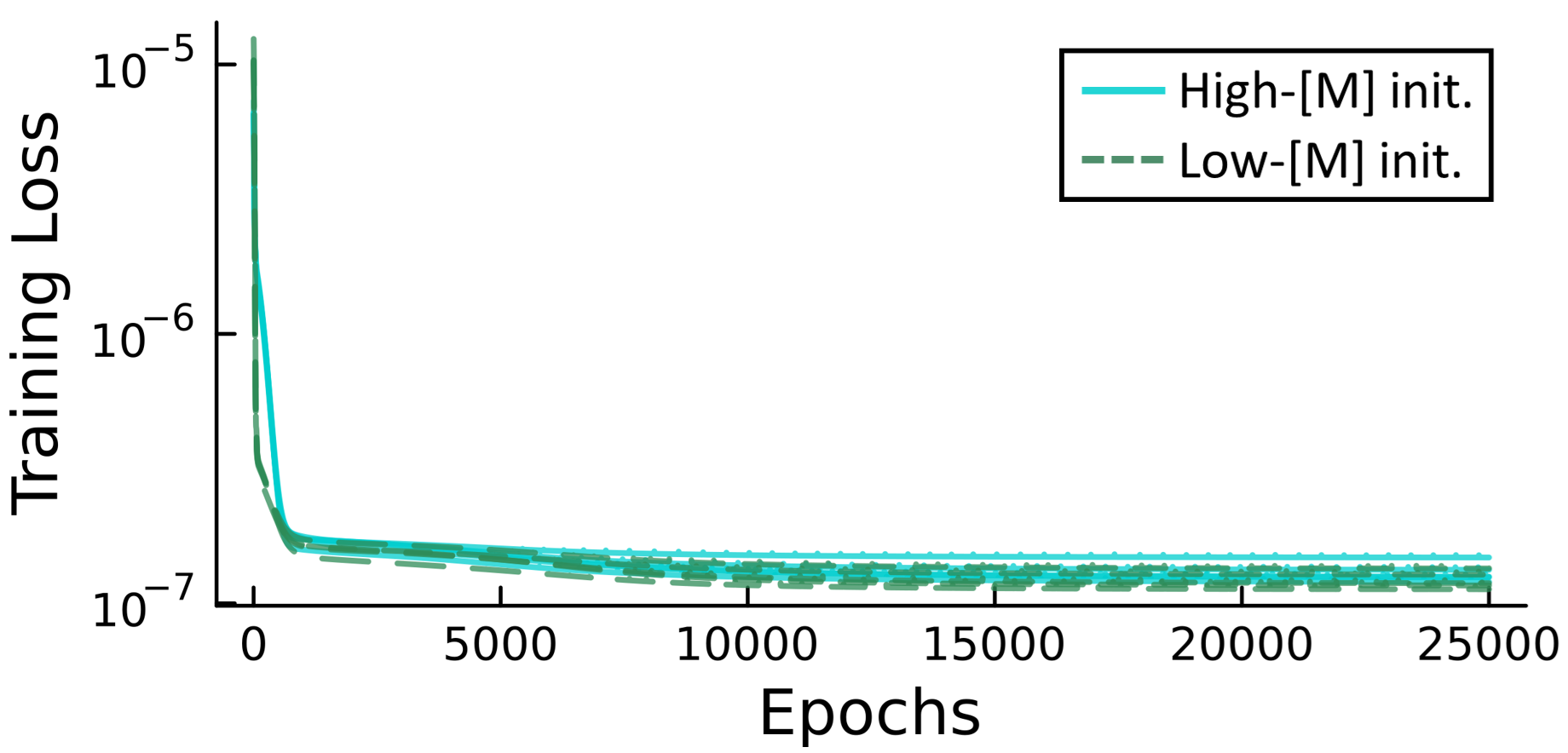}
	\caption{{Loss profiles of ten random initializations of the KA-CRNN on ten samples of the uncertain data: five sampled near the high-pressure model, five sampled near the low-pressure model. Rapid, stable, and initialization-invariant convergence seen in all cases.}}
	\label{fig:sec3:loss}
\end{figure}

\begin{figure}
\centering
\includegraphics[width=0.9\linewidth]{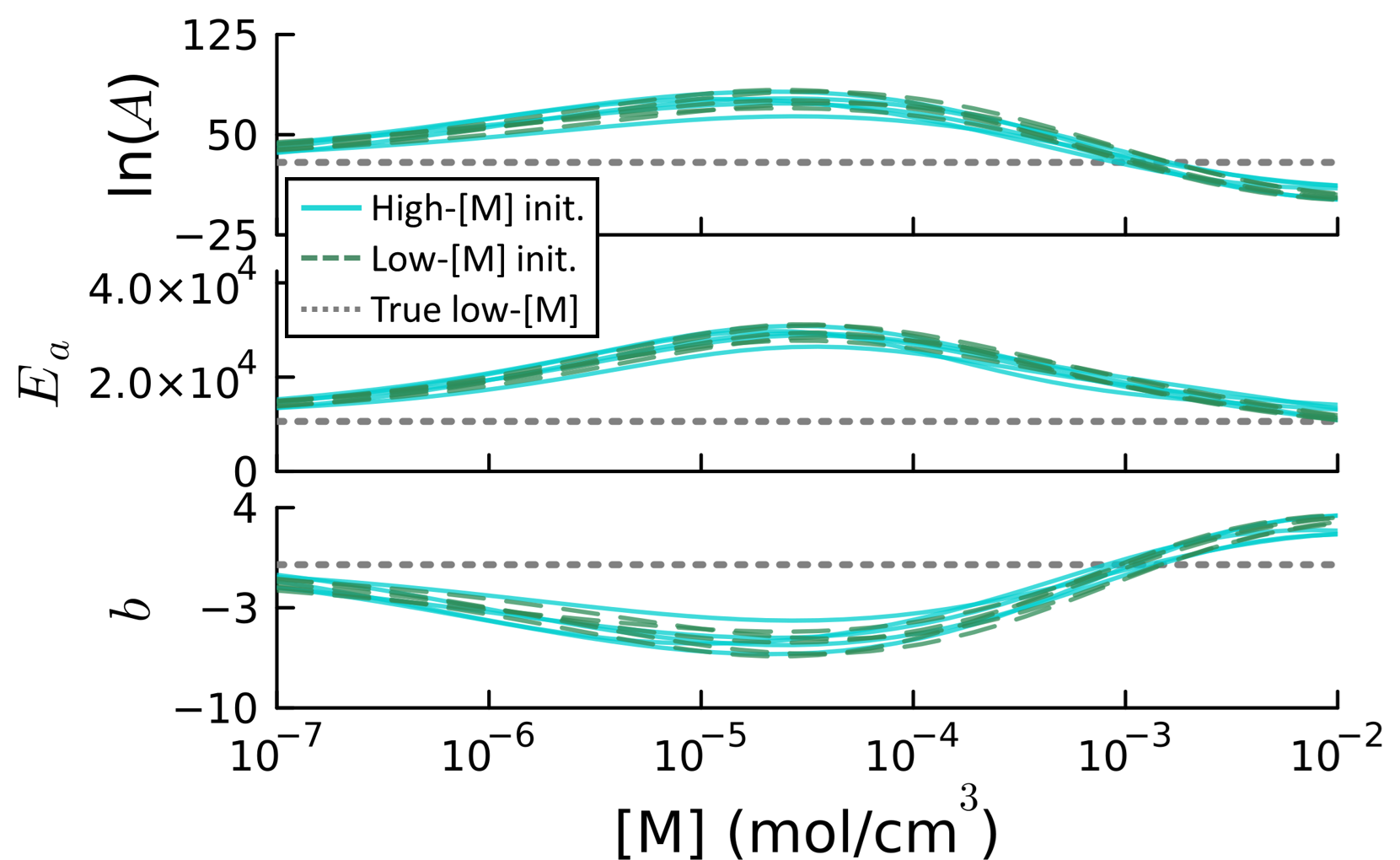}
	\caption{{KA-CRNN activations for all ten random parameter initializations, trained on ten different samples of uncertain data. Robust convergence seen across all cases, with some expected variation due to the noise contained in the data. Low-[M] parameters, shown as horizontal dashed grey lines, identified well in all cases.}}
	\label{fig:sec3:acts}
\end{figure}

{This inference process was carried out independently ten times: five times with random initializations near the high-pressure limit in the data ($[M]=10^{-2}$), and five times with random initializations near the global low-pressure limit (inferred here from the $[M]=10^{-7}$ data). Fits to the data in all cases are visually similar to the fit shown in Fig. {\ref{fig:sec3_recon}}. To demonstrate this robustness, we report the loss function dynamics for all ten initialization in Fig. {\ref{fig:sec3:loss}} and the learned activations for all ten initializations in Fig. {\ref{fig:sec3:acts}}. There, we see that regardless of noise in the data or randomly selected initialization on either end of the studied pressure regime, the KA-CRNN converges smoothly and robustly to a very similar model. The 25,000 training epochs used here, at 50 seconds per 1000 epochs, took on average 20 minutes to converge on a single 2018 Intel i9-9980XE CPU thread. Thus, the ten studied models took 20 minutes total to converge on ten threads. 25,000 epochs was roughly chosen based on when the model was no longer reducing its loss by 1$\%$ per 1,000 epochs, although in this case it seems that fewer epochs would also suffice. If needed, further speedup via code optimization, parallelization, or newer equipment is possible.}

\subsection{Future outlook, extension to different regimes}

As we turn to the discussion of potential future applications and further development, we identify two key pieces of the KA-CRNN modeling approach that may initially appear to be limitations: the linear mixture rule, and the falloff model form of the training data. We discuss here the simple model form changes that can be used to extend KA-CRNNs to more universal forms, some of which has in fact already been demonstrated in applications in the literature.

We reiterate first that the linear mixture rule was imposed here for convenience and to provide a simple comparison against a common approach in the literature - it is not built into the KA-CRNN model, and in fact, it is evaluated after the forward KA-CRNN gradient pass. Thus, in other situations, practitioners might choose to employ more complex nonlinear mixture rules like the Burke rate expressions {(LMR-R)} \cite{burke2017evaluating, singal2024implementation}, while retaining full swap-in compatibility with the KA-CRNN approach. {While not in the scope of the current work, we briefly discuss here that a major component of the LMR-R model is the use of a reduced mixture pressure R computed based on each individual collider's concentration and high/low pressure limit kinetics. In a KA-CRNN implementation, the $\text{KAN}\left(p_{ext}, \bm{\theta}_i\right)$ formulation in Eq. {\ref{eq:activ}}, where $p_{ext}$ is the collider-specific pressure, could be rewritten as $\text{KAN}\left(R, \bm{\theta}_i\right)$, either using the direct Burke model for $R$ or separately inferring $R$ as a function of temperature, pressure, and concentrations.}

Further, while we leveraged the empirical {Troe and SRI kinetic model forms} in this work to generate synthetic training data, the KA-CRNN is a fully general, [M]-sensitive Arrhenius-based approach with no further empirical assumptions built in: the falloff behavior learned in this work reflects the KA-CRNN's general fitting capability in a case with empirical falloff-based training data, rather than a presumed pressure model form. Thus, KA-CRNNs are trivially capable of being applied in cases with training data governed by other generalized kinetic rate forms, or even data where no known form exists. In fact, they are not structurally limited to [M]-sensitive kinetics at all, and has a proven ability to learn Arrhenius kinetics as an arbitrary function of any external quantity: a recent application of KA-CRNNs repurposed them from learning combustion kinetics as a function of pressure to learning battery failure kinetics as a function of the state of charge (SOC) \cite{koenig2025learning}. This extension leveraged the basic KA-CRNN algorithm with a fundamentally different input to deliver working models and novel scientific conclusions in a different research area. Its general, assumption-free $p_{ext}$ form is an all-purpose method to capture kinetic dependencies on [M], SOC, or other arbitrary external variables for which widely-accepted empirical forms do not exist or may not provide sufficient accuracy.

{Finally, we acknowledge that many combustion kinetic models are substantially larger than the targeted proofs of concept shown here, and provide some brief discussion suggesting the scalability of CRNNs. The original Neural ODE work {\cite{chen_neural_2019}} recognized scaling issues with forward-mode autodifferentiation and proposed use of the adjoint sensitivity approach for large ODE systems. While we leveraged the forward-mode approach here due to the compact size of the system (as is common with CRNNs {\cite{ji_autonomous_2022}}), the adjoint method (as also used with CRNNs {\cite{ji_autonomous_2021}}) may prove suitable with larger systems. Both approaches have been demonstrated to work with KAN-ODEs} \cite{koenig_kan-odes_2024, koenig_chemkans_2025}. {Thus, there is significant precedent in the literature for scaling CRNNs and KAN-ODEs (the two core building blocks of KA-CRNNs) to larger systems. In fact, CRNNs have been previously applied to combustion models with 30-40 species and 100-200 reactions {\cite{su2023kinetics}}, and even the previously discussed SOC application of the brand new KA-CRNNs was on a multi-species, multi-reactant system {\cite{koenig2025learning}}.}

\section{Conclusions\label{sec:Conclusions}} \addvspace{10pt}

This work proposes KA-CRNNs, a novel machine learning tool for automatic kinetic inference of reaction models subject to external factors. KA-CRNN's combination of an {Arrhenius-encoded CRNN backbone for temperature sensitivity with data-driven yet interpretable KAN pressure activations} enables accurate inference of kinetic rates with unknown dependencies on external parameters, like the system pressure or individual collider concentrations. Once trained, the KA-CRNN is fully visualizable and interpretable, enabling the elucidation and symbolic expression of external parameter effects. 

{An initial study moved beyond traditional CRNN capabilities by learning a pressure-sensitive third-body reaction. The learned Arrhenius parameters converged well to the underlying Troe high-pressure limit, suggesting true extraction of meaningful underlying physics. Comparison against a PLOG interpolant demonstrated its improved accuracy thanks to its elimination of the loglinear assumption. A deeper dive into mixture rules next showed the KA-CRNN inferring collider-specific models from confounding bath gas blends and randomized temperatures, with an unseen testing case demonstrating little to no overfitting. Overall trends in collider concentrations were properly extracted even when given the ``wrong'' form of the Arrhenius model, demonstrating convergence even in cases with little to no prior knowledge. Finally, the robustness of the algorithm was demonstrated via consistent convergence across a series of training expeirments: KA-CRNNs with randomized initializations were trained  on noise-augmented data generated from an SRI-parameterized reaction with an inverted pressure sensitivity.}

The KA-CRNN framework shows promise in combustion applications thanks to its capability to accurately, efficiently, and interpretably learn pressure-sensitive kinetics. It is also promising in arbitrary kinetic modeling problems where the effects of other external forcings on kinetic rates are studied, especially when current models do not exist or contain unknown parameters. In all applications, the visualizable activations {with interpretable trends in pressure} show significant promise for physical model development.

\section*{Acknowledgments} \addvspace{10pt}

The work is supported by the National Science Foundation (NSF) under Grant No. CBET-2143625. BCK is partially supported by the NSF Graduate Research Fellowship under Grant No. 1745302. The authors would like to thank Dr. Sunkyu Shin for valuable discussion regarding mixture rules.

% -------------------------------------------------------------------- %
% -------------------------------------------------------------------- %
% -------------------------------------------------------------------- %

% -------------------------------------------------------------------- %
% -------------------------------------------------------------------- %
% -------------------------------------------------------------------- %
\bibliography{ka-crnn}

% -------------------------------------------------------------------- %
% -------------------------------------------------------------------- %
% -------------------------------------------------------------------- %

\newpage

\small
\baselineskip 10pt

% -------------------------------------------------------------------- %
% -------------------------------------------------------------------- %
% -------------------------------------------------------------------- %

\end{document}